# Delayed Feedback Capacity of Stationary Sources over Linear Gaussian Noise Channels

Shaohua Yang
Marvell Semiconductor Inc
Santa Clara, CA 95054

Aleksandar Kavčić
University of Hawaii
Honolulu, HI 96822

*Abstract*—We consider a linear Gaussian noise channel used with delayed feedback. The channel noise is assumed to be a ARMA (autoregressive and/or moving average) process. We reformulate the Gaussian noise channel into an intersymbol interference channel with white noise, and show that the delayed-feedback of the original channel is equivalent to the instantaneous-feedback of the derived channel. By generalizing results previously developed for Gaussian channels with instantaneous feedback and applying them to the derived intersymbol interference channel, we show that conditioned on the delayed feedback, a conditional Gauss-Markov source achieves the feedback capacity and its Markov memory length is determined by the noise spectral order and the feedback delay. A Kalman-Bucy filter is shown to be optimal for processing the feedback. The maximal information rate for stationary sources is derived in terms of channel input power constraint and the steady state solution of the Riccati equation of the Kalman-Bucy filter used in the feedback loop.

## I. INTRODUCTION

For Gaussian noise channels used with feedback, the channel capacity has been characterized in various aspects. For memoryless channels, Shannon [1] showed that feedback does not increase the capacity, and Schalkwijk and Kalaith [2] proposed a capacity achieving feedback code. For channels with memory, bounds have been developed for the feedback capacity [3], [4], [5], [6], [7]. In [8], the optimal feedback source distribution is derived in terms of a state-space channel representation and Kalman filtering. The maximal information rate for stationary sources is derived in an analytically explicit form in [9]. For first order moving-average (MA) Gaussian noise channels, the feedback capacity is achieved by stationary sources as shown in [10].

Here we consider a Gaussian noise channel used with *delayed* feedback under an average-input-power constraint. Compared to the instantaneous feedback case, fewer results have been obtained on channels with delayed feedback. Yanagi [11] derived an upper bound on the finite block length delayed feedback capacity. In [12], it was shown that delayed feedback capacity for finite-state machine channels can be determined based on a method developed for instantaneous feedback by augmenting the channel state to account for feedback delay.

We first re-formulate the Gaussian noise channel with delayed feedback into an equivalent state-space channel model with *instantaneous* feedback and white noise. The delayed-feedback information rate of the original Gaussian noise channel equals the instantaneous-feedback information rate of the derived state-space channel. By generalizing the methodology and results derived in [9], [8], we show that

1) a feedback-dependent Gauss-Markov source is optimal for achieving the delayed-feedback capacity, and the necessary Markov memory length equals the larger of
   a) the moving average (MA) noise spectral order, and
   b) the sum of the feedback delay and the autoregressive (AR) noise spectral order;
2) a state estimator (Kalman-Bucy filter) for the derived state-space channel model is optimal for processing the (delayed) feedback information, and the solution of its steady-state Riccati equation delivers the maximal information rate for stationary sources.

**Notation:** Random variables are denoted by upper-case letters, e.g., $X_t$, and their realizations are denoted using lower case letters, e.g., $x_t$. A sequence $x_i, x_{i+1}, \ldots, x_j$ is shortly denoted by $x_i^j$. The letter E stands for the expectation. The differential entropy of a random variable $X$ is denoted by $h(X)$. Bold uppercase letters stand for matrices (e.g., $\mathbf{K}$), while underlined letters stand for column vectors (e.g., $\underline{c}$).

## II. CHANNEL MODEL REFORMULATION

Let $X_t$ be channel input at time $t$. Let $R_t$ be channel output at time $t$. We start by considering a Gaussian noise channel

$$R_t = X_t + N_t. \qquad (1)$$

The noise $N_t$ is assumed to be an autoregressive moving average (ARMA) random Gaussian process with a rational power spectrum

$$S_N(\omega) = \sigma_W^2 \frac{\left(1 - \sum_{m=1}^{M} a_m e^{-jm\omega}\right)\left(1 - \sum_{m=1}^{M} a_m e^{jm\omega}\right)}{\left(1 + \sum_{k=1}^{K} c_k e^{-jk\omega}\right)\left(1 + \sum_{k=1}^{K} c_k e^{jk\omega}\right)}. \qquad (2)$$

The coefficients $a_m$ and $c_k$ are the spectral poles and zeros, and $M$ and $K$ indicate the orders of the the moving-average (MA) and autoregressive (AR) noise power spectral components, respectively. Since the poles and zeros of (2) appear in pairs symmetric with respect to the unit circle [13], without loss of generality, we may assume that $|a_m| < 1$ and $|c_k| < 1$. Hence, the filter defined by

$$H(z) = \left(1 - \sum_{m=1}^{M} a_m z^{-m}\right) \Big/ \left(1 + \sum_{k=1}^{K} c_k z^{-k}\right) \qquad (3)$$

and its inverse are both causal, stable and invertible.

We make the following assumptions on the channel usage:
1) The power of the channel input process is constrained[1] $\lim_{n\to\infty} \mathrm{E}\left[\sum_{t=1}^{n} X_t^2\right]/n = P$.
2) Let $\nu > 1$ be the feedback delay. The prior channel outputs $R_{-\infty}^{t-\nu}$ are known to the transmitter (via the feedback loop) before the transmission of $X_t$.
3) Transmission starts at time $t = 1$, i.e., $X_t = 0$ for $t \leq 0$. Thus, noise history $N_{-\infty}^{-\nu}$ is known to both the transmitter and receiver.

Since the filter $H(z)$ is invertible, we may apply $H^{-1}(z)$ to the channel output $R_t$ without changing the channel capacity. The equivalent intersymbol interference (ISI) channel has $X_t$ as the channel input, $U_t$ as the channel output, and white Gaussian noise $V_t$ (with power $\sigma_W^2$).

$$U(z) = H^{-1}(z)\left(X(z) + N(z)\right) = H^{-1}(z)X(z) + V(z). \quad (4)$$

The original channel outputs $R_{-\infty}^t$ can be determined from $U_{-\infty}^t$ using filter $H(z)$.

To simplify notation for deriving the delayed information rate, we change variables $Y_t \triangleq U_{t-\nu}$ and $W_t = V_{t-\nu}$, and further reformulate the ISI channel in terms of $X_t$ and $Y_t$ as

$$Y(z) = z^{-\nu}U(z) = z^{-\nu}H^{-1}(z)X(z) + W(z). \quad (5)$$

The ISI channel (5) with input $X_t$ and output $Y_t = U_{t-\nu}$ is depicted in Fig 1. The channel is completely characterized by the tap coefficients $a_m$, $c_k$ and $\nu$. Without of loss of generality, we can assume[2] $M = K + \nu$, and denote

$$\underline{c} \triangleq \left[\underbrace{0,\cdots,0}_{\nu-1 \text{ zeros}}, 1, c_1, c_2, \cdots, c_K\right]^\mathrm{T}. \quad (6)$$

The channel depicted in Figure 1 has a state-space representation. Let the vector of values stored in the channel memory, i.e., $\underline{S}_t \triangleq [S_t(1), S_t(2), \ldots, S_t(M)]^\mathrm{T}$, be the channel *state* vector. The state space channel equations are

$$\underline{S}_t = \mathbf{A}\underline{S}_{t-1} + \underline{b}X_t \quad (7)$$
$$Y_t = \underline{c}^\mathrm{T}\underline{S}_{t-1} + W_t, \quad (8)$$

where $W_t$ is white Gaussian noise with variance $\sigma_W^2$. The constant square matrix $\mathbf{A}$ and vector $\underline{b}$ are defined as

$$\mathbf{A} \triangleq \begin{bmatrix} a_1 & a_2 & \ldots & a_{M-1} & a_M \\ 1 & 0 & \ldots & 0 & 0 \\ 0 & 1 & \ldots & 0 & 0 \\ \vdots & \vdots & \ddots & \vdots & \vdots \\ 0 & 0 & \ldots & 1 & 0 \end{bmatrix}, \quad \underline{b} \triangleq \begin{bmatrix} 1 \\ 0 \\ 0 \\ \vdots \\ 0 \end{bmatrix}. \quad (9)$$

From channel assumptions 1)-3), we have the following:

---

[1]Since it has been shown [14] that the feedback capacity is a concave function of $P$, it is not necessary to consider the inequality constraint $\lim_{n\to\infty} \mathrm{E}\left[\sum_{t=1}^{n} X_t^2\right]/n \leq P$.

[2]If $M < K + \nu$, we let $a_{M+1} = 0, \cdots, a_{K+\nu} = 0$ and then redefine $M$ to be $K + \nu$. If $M > K + \nu$, we let $c_{K+\nu+1} = 0, \cdots, c_{M-\nu} = 0$ and then redefine $K$ to be $M - \nu$.

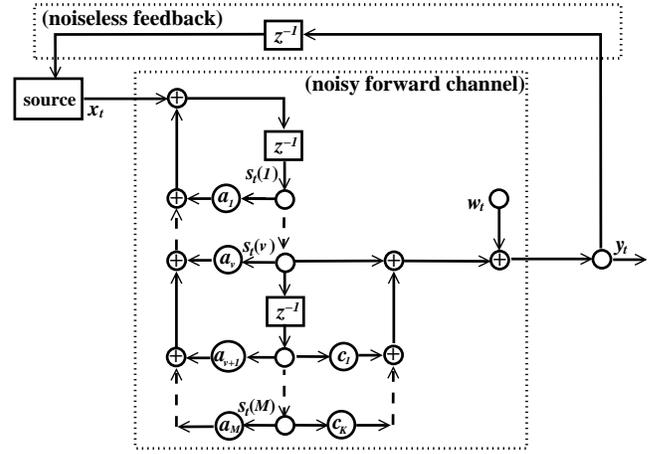

Fig. 1. Equivalent state space model for Gaussian channels with delayed feedback.

I) Since $X_t = 0$ for $t \leq 0$, the initial channel state $\underline{s}_0 = \underline{0}$ is known to both the transmitter and the receiver.
II) The sequences $\underline{S}_1^t$ and $X_1^t$ determine each other uniquely according to equation (7).
III) Given the channel state $\underline{S}_{t-1} = \underline{S}_{t-1}$, the channel output $Y_t$ is statistically independent of channel states $\underline{S}_0^{t-2}, \underline{S}_t$ and outputs $Y_1^{t-1}$, that is

$$P_{Y_t|\underline{S}_0^t, Y_1^{t-1}}\left(y_t \,|\, \underline{s}_0^t, y_1^{t-1}\right) = P_{Y_t|\underline{S}_{t-1}}\left(y_t \,|\, \underline{s}_{t-1}\right). \quad (10)$$

Since the variance of the process $W_t$ is $\sigma_W^2$, the conditional differential entropy of the channel output equals

$$h\left(Y_t \,|\, \underline{S}_0^t, Y_1^{t-1}\right) = h\left(Y_t \,|\, \underline{S}_{t-1}\right) = \frac{1}{2}\log(2\pi e\sigma_W^2). \quad (11)$$

IV) The instantaneous feedback of $Y_t$ in the above derived channel is equivalent to the delayed feedback $R_{t-\nu}$ in the original channel. Thus, we only need to consider the following encoder $X_t = X\left(\mathcal{M}, Y_1^{t-1}\right)$, where $\mathcal{M}$ is the message to transmit. For the source distribution, the channel input $X_t$ is causally dependent on all previous channel states $\underline{S}_0^{t-1}$ and channel outputs $Y_1^{t-1}$

$$P_t\left(x_t \,|\, \underline{s}_0^{t-1}, y_1^{t-1}\right) \triangleq P_{X_t|\underline{S}_0^{t-1}, Y_1^{t-1}}\left(x_t \,|\, \underline{s}_0^{t-1}, y_1^{t-1}\right), (12)$$

or equivalently in terms of the channel states as

$$P_t\left(\underline{s}_t \,|\, \underline{s}_0^{t-1}, y_1^{t-1}\right) \triangleq P_{\underline{S}_t|\underline{S}_0^{t-1}, Y_1^{t-1}}\left(\underline{s}_t \,|\, \underline{s}_0^{t-1}, y_1^{t-1}\right). (13)$$

We only need to consider Gaussian sources [4].

## III. INFORMATION RATE AND OPTIMAL SOURCES

We note that in the derived state-space channel model, the first channel output that carries non-zero signal is $Y_{1+\nu} = U_1$.

The information rate equals

$$\mathcal{I}(\mathcal{M};Y) \triangleq \lim_{n\to\infty} \frac{1}{n-\nu} I(\mathcal{M};Y_1^n | \underline{s}_0) \quad (14)$$

$$= \lim_{n\to\infty} \frac{1}{n} I(\mathcal{M};Y_1^n | \underline{s}_0) \quad (15)$$

$$= \lim_{n\to\infty} \frac{1}{n} [h(Y_1^n | \underline{s}_0) - h(W_1^n)] \quad (16)$$

$$= \lim_{n\to\infty} \frac{1}{n} h(Y_1^n | \underline{s}_0) - \frac{1}{2}\log(2\pi\sigma_W^2), \quad (17)$$

$$= \lim_{n\to\infty} \frac{1}{n} \sum_{t=1}^n [h(Y_t | \underline{s}_0, Y_1^{t-1})$$
$$- h(Y_t | \underline{s}_0, Y_1^{t-1}, S_{t-1})] \quad (18)$$

$$= \lim_{n\to\infty} \frac{1}{n} \sum_{t=1}^n [I(S_{t-1}, Y_t | \underline{s}_0, Y_1^{t-1})]. \quad (19)$$

In the following analysis, we note that since the initial channel state $\underline{s}_0$ is known according to the channel assumption in Section II, for notational simplicity, we will not explicitly write the dependence on $\underline{s}_0$ when obvious.

We consider all feedback-dependent Gaussian sources defined in (12) or (13)

$$\mathcal{P} \triangleq \{P_t(\underline{s}_t | \underline{s}_0^{t-1}, y_1^{t-1}), t=1,2,\ldots\}, \quad (20)$$

and the channel input is subject to the input power constraint

$$\lim_{n\to\infty} \mathrm{E}\left[\frac{1}{n}\sum_{t=1}^n (X_t)^2\right] = P. \quad (21)$$

The following two theorems can be conveniently generalized from [8] where they were originally derived for Gaussian channels used with instantaneous feedback.

*Theorem 1 (Gauss-Markov Source are Optimal):* For the power-constrained linear Gaussian channel, a feedback-dependent Gauss-Markov source

$$\mathcal{P}^{\mathrm{GM}} \triangleq \{P_t(\underline{s}_t | \underline{s}_{t-1}, y_1^{t-1}), t=1,2,\ldots\} \quad (22)$$

achieves the delayed-feedback channel capacity. (proof in Appendix) □

By Theorem 1, without loss of optimality, in the sequel we only consider feedback-dependent Gauss-Markov sources as in (22).

*Definition 1:* We use $\alpha_t(\cdot)$ as shorthand notation for the posterior pdf of the channel state $\underline{S}_t$, that is

$$\alpha_t(\underline{\mu}) \triangleq P_{\underline{S}_t | \underline{S}_0, Y_1^t}(\underline{\mu} | \underline{s}_0, y_1^t), \quad (23)$$

which is Gaussian due to Gaussian channel inputs. □

For a feedback-dependent Gauss-Markov source $\mathcal{P}^{\mathrm{GM}}$, the functions $\alpha_t(\cdot)$ can be recursively computed as

$$\alpha_t(\underline{\mu}) = \frac{\int \alpha_{t-1}(\underline{v}) P_t(\underline{\mu}|\underline{v}, y_1^{t-1}) P_{Y_t|\underline{S}_{t-1},\underline{S}_t}(y_t|\underline{v},\underline{\mu}) d\underline{v}}{\iint \alpha_{t-1}(\underline{v}) P_t(\underline{u}|\underline{v}, y_1^{t-1}) P_{Y_t|\underline{S}_{t-1},\underline{S}_t}(y_t|\underline{v},\underline{u}) d\underline{u} d\underline{v}}. \quad (24)$$

The Gaussian function $\alpha_t(\cdot)$ is completely characterized by the conditional mean $\underline{m}_t$ (vector of dimention $M$) and conditional covariance matrix $\mathbf{K}_t$ (of dimension $M$ by $M$)

$$\underline{m}_t = \mathrm{E}\left[\underline{S}_t | \underline{s}_0, y_1^t\right], \quad (25)$$
$$\mathbf{K}_t = \mathrm{E}\left[(\underline{S}_t - \underline{m}_t)(\underline{S}_t - \underline{m}_t)^{\mathrm{T}} | \underline{s}_0, y_1^t\right]. \quad (26)$$

We note that the recursion (24) can be implemented by a Kalman-Bucy filter.

*Theorem 2:* For the power-constrained linear Gaussian channel, the delayed-feedback capacity is achieved by a feedback-dependent Gauss-Markov source $\mathcal{P}_\alpha^{\mathrm{GM}}$ defined as

$$\mathcal{P}_\alpha^{\mathrm{GM}} \triangleq \{P_t(\underline{s}_t | \underline{s}_{t-1}, \alpha_{t-1}(\cdot)), t=1,2,\ldots\}, \quad (27)$$

where the Markov transition probability depends only on the posterior distribution function of the derived channel state $\alpha_t(\cdot)$ instead of all prior channel outputs. (proof in Appendix) □

Theorem 2 suggests that, for the task of constructing the next signal to be transmitted, all the "knowledge" contained in the vector of prior channel outputs is captured by the posterior distribution $\alpha_{t-1}(\cdot)$ of the channel state.

By Theorem 1 and Theorem 2, we only need to consider a feedback-dependent Gauss-Markov source $\mathcal{P}_\alpha^{\mathrm{GM}}$ as defined in (27).

## IV. Feedback Capacity Computation

The delayed feedback capacity thus can be derived in a similar way as in [8], though the results slightly differ due to feedback delay.

### A. Source Parameterization

Without loss of generality, a feedback-dependent Gauss-Markov source $\mathcal{P}_\alpha^{\mathrm{GM}}$ can be expressed as

$$X_t = \underline{d}_t^{\mathrm{T}} \underline{S}_{t-1} + e_t Z_t + g_t, \quad (28)$$

where $Z_t$ is a Gaussian random variable with zero-mean and unit-variance and is independent of $Z_1^{t-1}$, $X_1^{t-1}$ and $Y_1^{t-1}$, and vector $\underline{d}_t$ is of length $M$. The coefficients $\underline{d}_t$, $e_t$ and $g_t$ are all dependent on the Gaussian pdf $\alpha_{t-1}(\cdot)$, or alternatively on its mean $\underline{m}_{t-1}$ and covariance matrix $\mathbf{K}_{t-1}$. The set of coefficients $\{\underline{d}_t, e_t, g_t\}$ completely determine the transition probabilities of the feedback-dependent Gauss-Markov source $\mathcal{P}_\alpha^{\mathrm{GM}}$ defined in (27).

*Lemma 1:* For the feedback-dependent Gauss-Markov source as parameterized in (28), we have

$$h(Y_t | \underline{s}_0, y_1^{t-1}) - \frac{1}{2}\log(2\pi e\sigma_W^2) = \frac{1}{2}\log\left(1 + \frac{\underline{c}^{\mathrm{T}}\mathbf{K}_{t-1}\underline{c}}{\sigma_W^2}\right), \quad (29)$$

and

$$\mathrm{E}[(X_t)^2 | \underline{s}_0, y_1^{t-1}] = \left(\underline{d}_t^{\mathrm{T}} \underline{m}_{t-1} + g_t\right)^2 + \underline{d}_t^{\mathrm{T}} \mathbf{K}_{t-1} \underline{d}_t + (e_t)^2, \quad (30)$$

where the values of $\underline{d}_t$, $e_t$, $g_t$ depend on $\underline{m}_{t-1}$ and $\mathbf{K}_{t-1}$. □

*Proof:* The first and second order moments of the channel input $X_t$ and output $Y_t$ can be computed as

$$\mathrm{E}\left[(X_t)^2 \big| \underline{s}_0, y_1^{t-1}\right] = \left(\underline{d}_t^\mathrm{T} \underline{m}_{t-1} + g_t\right)^2 + \underline{d}_t^\mathrm{T} \mathbf{K}_{t-1} \underline{d}_t + (e_t)^2 \quad (31)$$

$$\mathrm{E}\left[Y_t \big| \underline{s}_0, y_1^{t-1}\right] = \underline{c}^\mathrm{T} \underline{m}_{t-1} \quad (32)$$

$$\mathrm{E}\left[\left(Y_t - \mathrm{E}\left[Y_t \big| \underline{s}_0, y_1^{t-1}\right]\right)^2 \big| \underline{s}_0, y_1^{t-1}\right] = \underline{c}^\mathrm{T} \mathbf{K}_{t-1} \underline{c} + \sigma_W^2. \quad (33)$$

Conditioned on $\underline{s}_0$ and $y_1^{t-1}$, the variable $Y_t$ has a Gaussian distribution with variance (33), thus we obtain (29). ∎

*Lemma 2:* The parameters of the optimal feedback-dependent Gauss-Markov source must satisfy

$$g_t = -\underline{d}_t^\mathrm{T} \underline{m}_{t-1}. \quad (34)$$

*Proof:* By Lemma 1 and equation (17), the value of $g_t$ does not affect the information rate, but choosing $g_t$ as in (34) minimizes the average input power for given $\underline{d}_t$ and $e_t$.

We note that this essentially follows the center of gravity necessary condition for optimal sources as derived in [15]. ∎

### B. Feedback Capacity for Stationary Sources

*Definition 2 (Stationary sources):* A *stationary* feedback-dependent (Gauss-Markov) source is a source that induces stationary channel input and output processes. An *asymptotically stationary* feedback-dependent (Gauss-Markov) source, in its limit as $t \to \infty$, induces stationary channel input and output processes. □

*Lemma 3:* For a stationary (or asymptotically stationary) feedback-dependent Gauss-Markov source, the covariance matrix $\mathbf{K}_t$ and source coefficients $\underline{d}_t$ and $e_t$ converge, i.e.,

$$\lim_{t \to \infty} \mathbf{K}_t = \mathbf{K}, \qquad \lim_{t \to \infty} \underline{d}_t = \underline{d}, \qquad \lim_{t \to \infty} e_t = e. \quad (35)$$

Here, the matrix $\mathbf{K}$ satisfies the stationary Kalman-Bucy filter equation (the algebraic Riccati equation)

$$\mathbf{K} = \mathbf{Q} \mathbf{K} \mathbf{Q}^\mathrm{T} + \underline{b}\, \underline{b}^\mathrm{T} e^2 - \frac{\mathbf{Q} \mathbf{K} \underline{c}\, \underline{c}^\mathrm{T} \mathbf{K} \mathbf{Q}^\mathrm{T}}{\underline{c}^\mathrm{T} \mathbf{K} \underline{c} + \sigma_W^2}, \quad (36)$$

where the matrix $\mathbf{Q}$ is defined as $\mathbf{Q} \triangleq \mathbf{A} + \underline{b}\, \underline{d}^\mathrm{T}$. The instantaneous channel input power converges as

$$\lim_{t \to \infty} \mathrm{E}\left[(X_t)^2 \big| \underline{s}_0, y_1^{t-1}\right] = \underline{d}^\mathrm{T} \mathbf{K}_{t-1} \underline{d} + (e)^2. \quad (37)$$

□

*Proof:* Since the (asymptotically) stationary source induces, in its limit as $t \to \infty$, stationary channel input and output processes, by definition the Kalman-Bucy filter has a steady state, and thus the sequences $\mathbf{K}_t$, $\underline{d}_t$ and $e_t$ converge. The Riccati equation (36) is obtained as the stationary form of the covariance matrix of the Kalman-Bucy filter. The limit in (37) follows (30) and (35). ∎

*Theorem 3 (Feedback capacity for stationary sources):* For a power constrained Gaussian channel used with $\nu$-time delayed feedback, the maximal information rate for stationary sources equals

$$C_\nu^\mathrm{fb} = \max_{\underline{d}, e} \frac{1}{2} \log\left(1 + \frac{\underline{c}^\mathrm{T} \mathbf{K} \underline{c}}{\sigma_W^2}\right) \quad (38)$$

where the maximization in (38) is taken under constraints

$$\underline{d}^\mathrm{T} \mathbf{K} \underline{d} + e^2 = P \quad (39)$$

$$\mathbf{K} = \mathbf{Q} \mathbf{K} \mathbf{Q}^\mathrm{T} + \underline{b}\, \underline{b}^\mathrm{T} e^2 - \frac{\mathbf{Q} \mathbf{K} \underline{c}\, \underline{c}^\mathrm{T} \mathbf{K} \mathbf{Q}^\mathrm{T}}{\underline{c}^\mathrm{T} \mathbf{K} \underline{c} + \sigma_W^2}. \quad (40)$$

The matrix $\mathbf{Q}$ is defined as $\mathbf{Q} \triangleq \mathbf{A} + \underline{b}\, \underline{d}^\mathrm{T}$, and the matrix $\mathbf{K}$ is constrained to be non-negative definite. □

*Proof:* By Lemma 3, for any (asymptotically) stationary Gauss-Markov source, the sequences $\mathbf{K}_t$, $\underline{d}_t$ and $e_t$ converge as $t \to \infty$, so (17) and (29) turn into (38) as $n \to \infty$. Constraint (40) is the algebraic Riccati equation (36). Constraint (39) is the input power of the stationary source, and subsequently utilizing Lemmas 2 and 3. ∎

In general, the optimization problem in Theorem 3 involves $O(M^2)$ variables and can be conveniently solved analytically for small $M$ or numerically for large $M$.

## V. CONCLUSION

In this paper, we derived the delayed feedback capacity of power-constrained stationary sources over linear Gaussian channels with ARMA Gaussian noise. We first reformulated the linear Gaussian noise channel into a state-space form that is suitable for manipulating the delayed feedback information rate. Then, we obtained the delayed feedback capacity for stationary sources by generalizing and applying a method that was originally developed for computing the instantaneous feedback capacity. We showed that a feedback-dependent Gauss-Markov source achieves the delayed-feedback channel capacity and that the Kalman-Bucy filter is optimal for processing the feedback. The delayed-feedback capacity is expressible as an optimization problem with constraints on the conditional state covariance matrix of the Kalman-Bucy filter.

## APPENDIX

### A. Sketch of Proof for Theorem 1

Let $\mathcal{P}_1$ be any valid feedback-dependent Gaussian source distribution (not necessarily Markov) defined as

$$\mathcal{P}_1 \triangleq \left\{ P_t\left(\underline{s}_t \big| \underline{s}_0^{t-1}, y_1^{t-1}\right), t = 1, 2, \cdots \right\}. \quad (41)$$

From $\mathcal{P}_1$, we construct a Markov (not necessarily stationary) source distribution $\mathcal{P}_2$ as

$$\mathcal{P}_2 = \left\{ Q_t\left(\underline{s}_t \big| \underline{s}_{t-1}, y_1^{t-1}\right), t = 1, 2, \cdots \right\}. \quad (42)$$

where the functions $Q_t\left(\underline{s}_t \big| \underline{s}_{t-1}, y_1^{t-1}\right)$ are defined as the conditional marginal pdf's computed from $\mathcal{P}_1$

$$Q_t\left(\underline{s}_t \big| \underline{s}_{t-1}, y_1^{t-1}\right) \triangleq P^{(\mathcal{P}_1)}_{\underline{S}_t | \underline{S}_{t-1}, Y_1^{t-1}}\left(\underline{s}_t \big| \underline{s}_{t-1}, y_1^{t-1}\right) \quad (43)$$

We next show by induction that the the sources $\mathcal{P}_1$ and $\mathcal{P}_2$ induce the same distribution of $\underline{S}_{t-1}^t$ and $Y_1^t$, i.e.,

$$P^{(\mathcal{P}_1)}_{\underline{S}_{t-1}^t, Y_1^t | \underline{S}_0}\left(\underline{s}_{t-1}^t, y_1^t \big| \underline{s}_0\right) = P^{(\mathcal{P}_2)}_{\underline{S}_{t-1}^t, Y_1^t | \underline{S}_0}\left(\underline{s}_{t-1}^t, y_1^t \big| \underline{s}_0\right). \quad (44)$$

For $t = 1$, by the definition of source $\mathcal{P}_2$ we have

$$P^{(\mathcal{P}_2)}_{\underline{S}_1, Y_1 | \underline{S}_0}(\underline{s}_1, y_1 | \underline{s}_0) = Q_1(\underline{s}_1 | \underline{s}_0) P_{Y_1 | \underline{S}_0^1}(y_1 | \underline{s}_0^1) \quad (45)$$

$$= P_1(\underline{s}_1 | \underline{s}_0) P_{Y_1 | \underline{S}_0^1}(y_1 | \underline{s}_0^1) \quad (46)$$

$$= P^{(\mathcal{P}_1)}_{\underline{S}_1, Y_1 | \underline{S}_0}(\underline{s}_1, y_1 | \underline{s}_0). \quad (47)$$

Since $\underline{s}_0$ is known, this directly implies

$$P^{(\mathcal{P}_2)}_{\underline{S}_0^1, Y_1 | \underline{S}_0}(\underline{s}_0^1, y_1 | \underline{s}_0) = P^{(\mathcal{P}_1)}_{\underline{S}_0^1, Y_1 | \underline{S}_0}(\underline{s}_0^1, y_1 | \underline{s}_0). \quad (48)$$

Now, assume that the equality (44) holds for up to time $t - 1$, where $t > 1$, particularly,

$$P^{(\mathcal{P}_2)}_{\underline{S}_{t-2}^{t-1}, Y_1^{t-1} | \underline{S}_0}(\underline{s}_{t-2}^{t-1}, y_1^{t-1} | \underline{s}_0) = P^{(\mathcal{P}_1)}_{\underline{S}_{t-2}^{t-1}, Y_1^{t-1} | \underline{S}_0}(\underline{s}_{t-2}^{t-1}, y_1^{t-1} | \underline{s}_0) \quad (49)$$

$$= \int \prod_{\tau=1}^{t-1} P_\tau(\underline{s}_\tau | \underline{s}_0^{\tau-1}, y_1^{\tau-1}) P_{Y_\tau | \underline{S}_{\tau-1}^\tau}(y_\tau | \underline{s}_{\tau-1}^\tau) d\underline{s}_1^{t-3}. \quad (50)$$

The induction step for time $t$ is simply shown as follows

$$P^{(\mathcal{P}_2)}_{\underline{S}_{t-1}^t, Y_1^t | \underline{S}_0}(\underline{s}_{t-1}^t, y_1^t | \underline{s}_0)$$

$$= Q_t(\underline{s}_t | \underline{s}_{t-1}, y_1^{t-1}) \times P_{Y_t | \underline{S}_{t-1}^t}(y_t | \underline{s}_{t-1}^t) \times$$

$$\int P^{(\mathcal{P}_2)}_{\underline{S}_{t-2}^{t-1}, Y_1^{t-1} | \underline{S}_0}(\underline{s}_{t-2}^{t-1}, y_1^{t-1} | \underline{s}_0) d\underline{s}_{t-2} \quad (51)$$

$$\stackrel{(a)}{=} \frac{\int \left[\prod_{\tau=1}^{t-1} P_\tau(\underline{s}_\tau | \underline{s}_0^{\tau-1}, y_1^{\tau-1}) f_{Y_\tau | \underline{S}_{\tau-1}^\tau}(y_\tau | \underline{s}_{\tau-1}^\tau)\right] P_t(\underline{s}_t | \underline{s}_0^{t-1}, y_1^{t-1}) d\underline{s}_1^{t-2}}{\int \left[\prod_{\tau=1}^{t-1} P_\tau(\underline{s}_\tau | \underline{s}_0^{\tau-1}, y_1^{\tau-1}) f_{Y_\tau | \underline{S}_{\tau-1}^\tau}(y_\tau | \underline{s}_{\tau-1}^\tau)\right] d\underline{s}_1^{t-2}} \times$$

$$P_{Y_t | \underline{S}_{t-1}^t}(y_t | \underline{s}_{t-1}^t) \times \int \left[\prod_{\tau=1}^{t-1} P_\tau(\underline{s}_\tau | \underline{s}_0^{\tau-1}, y_1^{\tau-1}) f_{Y_\tau | \underline{S}_{\tau-1}^\tau}(y_\tau | \underline{s}_{\tau-1}^\tau)\right] d\underline{s}_1^{t-2} \quad (52)$$

$$\stackrel{(b)}{=} \int \prod_{\tau=1}^{t} P_\tau(\underline{s}_\tau | \underline{s}_0^{\tau-1}, y_1^{\tau-1}) P_{Y_\tau | \underline{S}_{\tau-1}^\tau}(y_\tau | \underline{s}_{\tau-1}^\tau) d\underline{s}_1^{t-2} \quad (53)$$

$$= P^{(\mathcal{P}_1)}_{\underline{S}_{t-1}^t, Y_1^t | \underline{S}_0}(\underline{s}_{t-1}^t, y_1^t | \underline{s}_0), \quad (54)$$

where $(a)$ is the result of expanding the definition in (43) for source $\mathcal{P}_2$ and the induction assumption (50) using the Bayes rule and substituting them into (51), and $(b)$ is obtained by simplifying the expression in (52).

Thus, we have shown that the channel states $\underline{S}_{t-1}^t$ and outputs $Y_1^t$ induced by sources $\mathcal{P}_1$ and $\mathcal{P}_2$ have the same distribution. It is therefore clear that the non-Markov source $\mathcal{P}_1$ and Markov source $\mathcal{P}_2$ induce the same information rate according to equality (19).

### B. Sketch of Proof for Theorem 2

Suppose that two different feedback vectors $\tilde{y}_1^{t-1}$ and $y_1^{t-1}$ ($\tilde{y}_1^{t-1} \neq y_1^{t-1}$) induce the same posterior channel state pdf $\alpha_{t-1}(\cdot)$, i.e., for any possible state value $\underline{s}_{t-1} = \underline{\mu}$ we have

$$P_{\underline{S}_{t-1} | \underline{S}_0, Y_1^{t-1}}(\underline{\mu} | \underline{s}_0, \tilde{y}_1^{t-1}) = P_{\underline{S}_{t-1} | \underline{S}_0, Y_1^{t-1}}(\underline{\mu} | \underline{s}_0, y_1^{t-1}). \quad (55)$$

Now consider two distributions for the source $S_\tau$, for $\tau \geq t$, the first distribution conditioned on $y_1^{t-1}$, and the second conditioned on $\tilde{y}_1^{t-1}$. If we let these two distributions be equal to each other for $\tau \geq t$, that is, if

$$\{P_\tau(\underline{s}_\tau | \underline{s}_{\tau-1}, \tilde{y}_1^{t-1}, y_t^{\tau-1}), \tau \geq t\}$$
$$= \{P_\tau(\underline{s}_\tau | \underline{s}_{\tau-1}, y_1^{t-1}, y_t^{\tau-1}), \tau \geq t\}, \quad (56)$$

then we have for any $k \geq t$

$$P_{Y_t^k, \underline{S}_{t-1}^k | \underline{S}_0, Y_1^{t-1}}(y_t^k, \underline{s}_{t-1}^k | \underline{s}_0, \tilde{y}_1^{t-1})$$

$$= \alpha_{t-1}(\underline{s}_{t-1}) \prod_{\tau=t}^{k} P_\tau(\underline{s}_\tau | \underline{s}_{\tau-1}, y_1^{\tau-1}) P_{Y_\tau | \underline{S}_{\tau-1}^\tau}(y_\tau | \underline{s}_{\tau-1}^\tau)$$

$$= P_{Y_t^k, \underline{S}_{t-1}^k | \underline{S}_0, Y_1^{t-1}}(y_t^k, \underline{s}_{t-1}^k | \underline{s}_0, y_1^{t-1}). \quad (57)$$

This shows that for any $k \geq t$ the entropies are equal

$$h(Y_t^k | \underline{s}_0, \tilde{y}_1^{t-1}) = h(Y_t^k | \underline{s}_0, y_1^{t-1}), \quad (58)$$

and for any $\tau \geq t$ the powers are equal

$$\mathrm{E}\left[(X_\tau)^2 | \underline{s}_0, \tilde{y}_1^{t-1}\right] = \mathrm{E}\left[(X_\tau)^2 | \underline{s}_0, y_1^{t-1}\right]. \quad (59)$$

Therefore, the optimal source distribution for time $\tau \geq t$ when $y_1^{t-1}$ is the feedback vector, must also be optimal when $\tilde{y}_1^{t-1}$ is the feedback vector, and vice versa. Since time $t$ is arbitrary, we conclude that, for any $t > 0$, the function $\alpha_{t-1}(\cdot)$ extracts from $y_1^{t-1}$ all that is necessary for formulating the optimal source distribution functions $P_t(\underline{s}_t | \underline{s}_{t-1}, y_1^{t-1})$.

### REFERENCES

[1] C. E. Shannon, "The zero error capacity of a noisy channel," *IRE Trans. on Inform. Theory*, vol. IT-2, pp. 112–124, Sept. 1956.
[2] J. Schalkwijk and T. Kailath, "A coding scheme for additive noise channels with feedback–I: No bandwidth constraint," *IEEE Transactions on Information Theory*, vol. 12, pp. 172–182, Apr. 1966.
[3] S. Butman, "A general formulation of linear feedback communications systems with solutions," *IEEE Trans. Inform. Theory*, vol. 15, pp. 392–400, May 1969.
[4] T. M. Cover and S. Pombra, "Gaussian feedback capacity," *IEEE Trans. Inform. Theory*, vol. 35, pp. 37–43, Jan 1989.
[5] M. Pinsker, "Talk delivered at the Soviet Information Theory Meeting." unpublished, 1969.
[6] P. Ebert, "The capacity of the Gaussian channel with feedback," *Bell Syst. Tech. J.*, vol. 49, pp. 1705–1712, Oct 1970.
[7] L. H. Ozarow, "Upper bounds on the capacity of Gaussian channels with feedback," *IEEE Trans. Inform. Theory*, vol. 36, pp. 156–161, Jan 1990.
[8] S. Yang, A. Kavčić, and S. Tatikonda, "On the feedback capacity of power constrained Gaussian noise channels with memory." Accepted for publication by *IEEE Transactions on Information Theory*.
[9] S. Yang, A. Kavčić, and S. Tatikonda, "Feedback capacity of stationary sources over Gaussian intersymbol interference channels," in *Proc. IEEE GLOBECOM 2006*, (San Francisco, USA), Nov 2006.
[10] Y.-H. Kim, "Feedback capacity of the first-order moving average gaussian channel," in *Proc. IEEE ISIT*, (Adelaide, Australia).
[11] K. Yanagi, "On the capacity of the discrete-time Gaussian channel with delayed feedback," *IEEE Transactions on Information Theory*, vol. 41, pp. 1051–1059, Jul. 1995.
[12] S. Yang, A. Kavčić, and S. Tatikonda, "The feedback capacity of finite-state machine channels," *IEEE Transactions on Information Theory*, vol. 51, pp. 799–810, Mar. 2005.
[13] A. V. Oppenheim and R. W. Schafer, *Discrete-Time Signal Processing*. Englewoods Cliffs, NJ: Prentice Hall, 1989.
[14] K. Yanagi, H. W. Chen, and J. W. Yu, "Operator inequality and its application to capacity of Gaussian channel," *Taiwanese Journal of Mathematics*, vol. 4, Sept. 2000.
[15] J. Schalkwijk, "Center-of-gravity information feedback," *IEEE Transactions on Information Theory*, vol. 14, pp. 324–331, Mar. 1968.